# AmpliNetECG12: A lightweight SoftMax-based relativistic amplitude amplification architecture for 12 lead ECG classification.

**Shreya Srivastava**
*Indian Institute of Technology*
*Roorkee, Roorkee, India*
ssrivastava@bt.iitr.ac.in

## ABSTRACT

The urgent need to promptly detect cardiac disorders from 12-lead Electrocardiograms using limited computations is motivated by the heart's fast and complex electrical activity and restricted computational power of portable devices. Timely and precise diagnoses are crucial since delays might significantly impact patient health outcomes. This research presents a novel deep-learning architecture that aims to diagnose heart abnormalities quickly and accurately. We devised a new activation function called ***aSoftMax***, designed to improve the visibility of ECG deflections. The proposed activation function is used with Convolutional Neural Network architecture to includes kernel weight sharing across the ECG's various leads. This innovative method thoroughly generalizes the global 12-lead ECG features and minimizes the model's complexity by decreasing the trainable parameters. aSoftMax, combined with enhanced CNN architecture yielded ***AmpliNetECG12***, we obtain exceptional accuracy of 84% in diagnosing cardiac disorders. AmpliNetECG12 shows outstanding prediction ability when used with the CPSC2018 dataset for arrhythmia classification. The model attains an F1-score of 80.71% and a ROC-AUC score of 96.00%, with 280,000 trainable parameters which signifies the lightweight yet efficient nature of AmpliNetECG12. The stochastic characteristics of aSoftMax, a fundamental element of AmpliNetECG12, improve prediction accuracy and also increasse the model's interpretability. This feature enhances comprehension of important ECG segments in different forms of arrhythmias, establishing a new standard of explainable architecture for cardiac disorder classification.

# HIGHLIGHTS

- AmpliNetECG12 is a novel architecture for processing long-term 12-lead ECG as a whole.
- It is a new activation function for relativistic amplification of ECG deflection amplitudes.
- Remarkable accuracy of 84% and F1-Score of 81% using 280k parameters and 12 lead ECG of just 50Hz.

# GRAPHICAL ABSTRACT

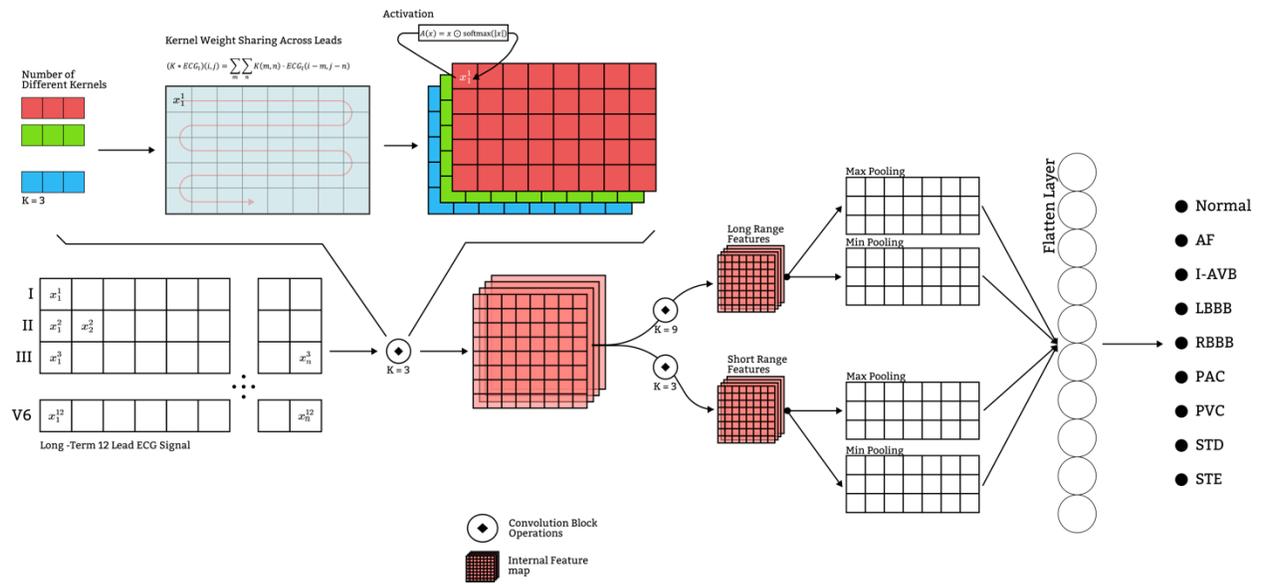

## 1. Introduction

Heart diseases and strokes have become much more common around the world, almost doubling from 271 million cases in 1990 to 523 million cases in 2019. In the same way, cardiovascular diseases have caused more deaths, from 12.1 million in 1990 to 20.5 million in 2021. CVDs are now becoming more common and killing more people [1]. This has put a lot of stress on healthcare systems, especially in developing countries where CVDs are more common and rising faster. The fact that healthcare services in these places are hard to afford makes this problem even worse. Because of these past patterns, it is now more important than ever to detect CVDs quickly and correctly. Electrocardiograms are one of the most important diagnostic tools for heart diseases [2]. This technology, which is painless and doesn't cost much, records the heart's electrical activity. It tells us the important things about heart rhythm, conduction patterns, and heart's health in general. ECGs (Figure 1) are being used for many things, like finding arrhythmias, myocardial infarctions, and changes in ischemia [3]. Because of this, it is important in both short-term and long-term medical situations. The ECG's ability to quickly and accurately pick up on changes in the heart's activity is very important for finding and treating CVDs early on. Physiologists use their years of training and experience to figure out what's wrong with the heart by reading complicated heart rhythms by hand. However, algorithms and machine learning are used in automated ECG analysis that uses artificial intelligence to look at these patterns quickly and correctly [4]. AI-based analysis can handle huge amounts of data, speed up diagnosis, and maybe even improve accuracy, especially when it comes to finding subtle or uncommon abnormalities. Manual analysis, on the other hand, requires more detailed knowledge and judgement based on the situation. Heart diseases are becoming more common, and there is a huge amount of ECG data that needs to be analyzed automatically. AI's inclusion into ECG analysis has been studied progressively because it might democratize healthcare, diagnose CVDs early, and help design individualized treatments.

Traditional methods for arrhythmia classification typically involve hand-extracted features from ECG signals and these features are then used to differentiate various types of arrhythmias by rule-based methods or machine learning methods for instance Z. Qibin *et al.* used Wavelet Transformation and Support Vector Machines [5], P. Tadejko *et al.* used Mathematical morphology transformations and Self-Organization Map and Learning Vector Quantization [6], A. Jovic *et al.* applied Chaos theory to ECG feature extraction [7], and S. Sahoo *et al.* detect QRS complex features based on the multiresolution wavelet transform [8]. These methods rely heavily on the expertise in signal processing and understanding of the ECG's physiological aspects, making it somewhat subjective and potentially variable in accuracy. In recent years these handcrafted features are further processed using machine learning algorithms most used algorithms are SVM, RF, KNN [9-11].

***Deep learning-based ECG analysis***: The transition to deep learning methods in ECG classification marks a significant shift from traditional feature engineering approaches. Deep learning algorithms have the ability to automatically learn and extract features from raw ECG data, eliminating the need for explicit feature engineering. Various neural networks, including CNN [12], RNN [13], LSTM [14], BiLSTM [15] and Transformer [16], have been used to identify arrhythmia. While scientists have also used combination of different Deep learning architecture and multiple data modalities for instance C. Chen *et al*. used CNN-LSTM [17], and Z. Ahmad *et al*. used Gramian Angular Field (GAF), Recurrence Plot (RP) and Markov Transition Field (MTF) of ECG signal as input to the CNN [18].

*12 lead ECG analysis*: While the above-mentioned approaches primarily tackle the problem of heartbeat (Smaller Section of ECG) classification into cardiac disorder classes. There is also a significant literature for long term ECG classification using publicly available datasets. Recent studies have explored multi-label classification of ECG signals using the public CPSC2018 dataset. Notable contributions include He et al.'s deep neural network combining a residual convolution module and a bidirectional LSTM with a high F1-score [15], Yao et al.'s attention-based time-incremental CNN for variable-length ECG signals [19], and Zhang et al.'s use of the Shapley Additive exPlanations method for model interpretation [20]. These studies highlight a shift from traditional methods, focusing on direct input of 12-lead ECG signals into neural networks and emphasizing leads with higher contributions for enhanced performance.

This research article presented a new method for cardiac diagnostics using deep learning. It involves the creation of a lightweight Convolutional based Neural Network architecture. The architecture is improved by including a unique activation function inspired from the Swish activation [21]. The novelty of this work lies in the capacity of proposed architecture to get almost state-of-the-art outcomes in ECG classification with just 280k trainable parameters and using compressed ECG data with a much lower sample frequency of just 50Hz. This signifies a decrease by a factor of 10 from the initial signal frequency often used in ECG analysis. An essential component of this study is the introduction of newly created activation function termed aSoftmax. The efficacy of aSoftmax activation for the proposed architecture is likely responsible for high accuracy even with lower resolution data. This has significant ramifications for practical healthcare uses, particularly in environments with limited resources where the capacity for storing and transmitting data may be restricted.

Proposed architecture uses the strategy of kernel weight sharing across given ECG leads, instead of the typical approach of training separate feature extraction models for each lead. This novel technique optimizes the model by enabling the use of identical kernels (or filters) in the convolutional layers across different leads. As a result, the model's trainable parameters are significantly reduced without any detrimental impact on its performance. Kernel sharing exploits the notion that certain inherent patterns and characteristics in ECG signals are universally present across several leads, hence eliminating the need for distinct feature extractors for each lead. This not only results in a smaller model with fewer parameters, but also has the potential to improve the model's capacity to generalize and be resistant to disturbances, as it acquires a more comprehensive representation of the heart's electrical activity. This strategy effectively tackles issues like as overfitting, computational efficiency, and resource consumption. It guarantees that the model maintains a low weight and is easier to handle, enabling quicker training and using less computing capacity. This is especially advantageous for implementing such models in practical situations, such as on portable/wearable devices and in distant places with restricted computational resources.

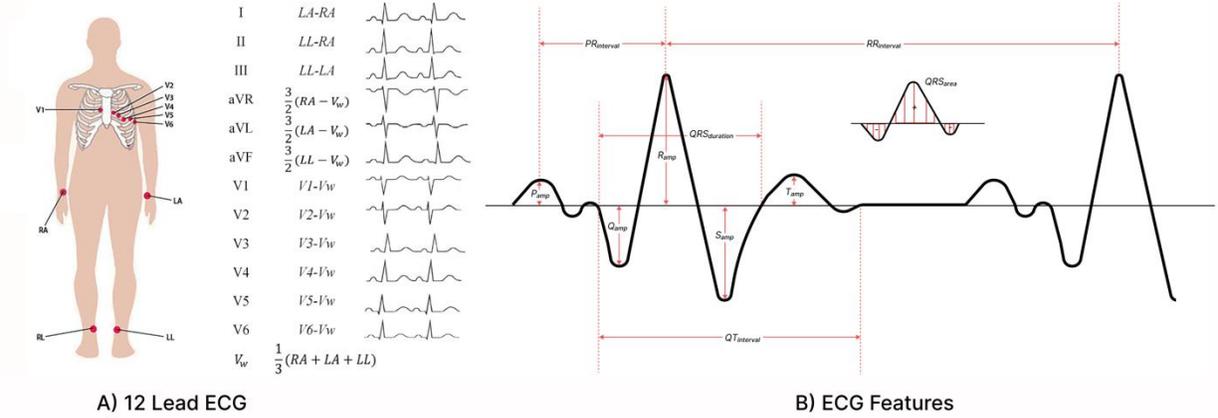

**Figure 1:** A) 12 Lead ECG collection system and B) ECG signal morphological and temporal features illustration.

**Table 1**
CPSC 2018 dataset description

| Class | Frequency (%) | Male (%) | Age | Time length (s) | | |
|---|---|---|---|---|---|---|
| | | | | Min | Max | Mean |
| **NSR** | 918 (13.35) | 363 (39.54) | 41.56 ± 18.45 | 10 | 60 | 15.43 |
| **AF** | 1221 (17.75) | 692 (56.67) | 71.47 ± 12.53 | 9 | 60 | 15.01 |
| **IAVB** | 722 (10.50) | 490 (67.87) | 66.97 ± 15.67 | 10 | 60 | 14.32 |
| **LBBB** | 236 (03.43) | 117 (49.58) | 70.48 ± 12.55 | 9 | 60 | 14.92 |
| **RBBB** | 1857 (27.00) | 1203 (64.78) | 62.84 ± 17.07 | 10 | 60 | 14.42 |
| **PAC** | 616 (08.96) | 328 (53.25) | 66.56 ± 17.71 | 9 | 60 | 19.46 |
| **PVC** | 700 (10.18) | 357 (51.00) | 58.37 ± 17.90 | 6 | 60 | 20.21 |
| **STD** | 869 (12.64) | 252 (29.00) | 54.61 ± 17.49 | 8 | 60 | 15.13 |
| **STE** | 220 (03.20) | 180 (81.82) | 52.32 ± 19.77 | 10 | 60 | 17.15 |

To Summarize, this research's main contributions are as follows:

1. We propose a new deep learning architecture specifically developed for long-term electrocardiogram (ECG) classification, characterized by fewer trainable parameters. This architectural design showcases improved efficacy in categorizing compressed electrocardiogram (ECG) data.
2. We propose a novel activation function, termed "aSoftmax activation," designed for transforming electrocardiogram

(ECG) signals. This function employs a relativistic amplitude amplification approach to accentuate the peaks in ECG data.
3. Our proposed approach has achieved notable results, attaining an accuracy of 84% and an F1-score of 0.81 on CPSC2018 dataset. This performance was achieved with a model comprising only 280,000 trainable parameters and while analyzing ECG data sampled at a frequency of just 50Hz.

## 2. Materials and methods
### 2.1 Dataset
The CPSC-2018 [22], provided as part of the 7th International Conference on Biomedical Engineering and Biotechnology, achieved a noteworthy accomplishment in ECG research by making a complete ECG dataset accessible to the public.

This dataset is a valuable resource for the worldwide research community. It consists of 6,877 12-lead ECG recordings, with an approx. equal number of individuals from both genders: 3,178 females and 3,699 males further statistical information about data is provided in Table 1. Each recording is obtained at a high sampling rate of 500 Hz and has a length ranging from 6 to 60 seconds. The amplitude of the recordings is measured in millivolts. The dataset is characterized by its

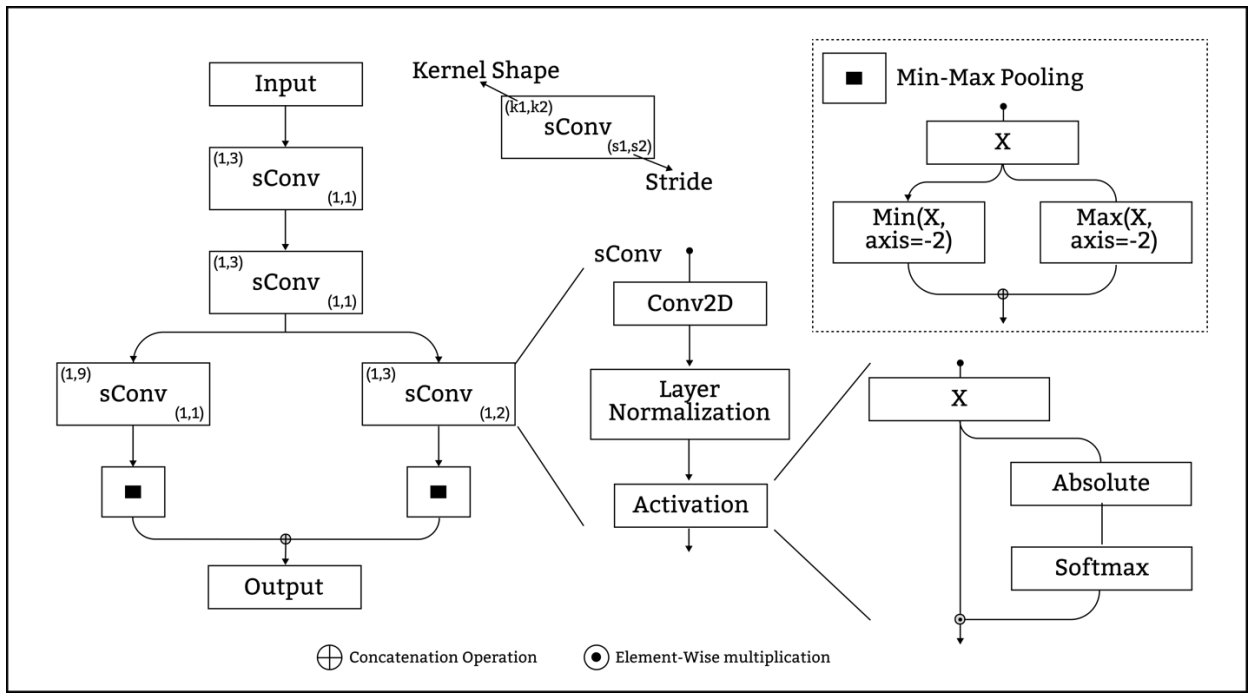

**Figure 2:** Graphical representation of proposed model

wide range and variety, which is emphasized by its multi-label categorization, including 9 diagnostic categories. The cardiac rhythms that are included are Normal Sinus Rhythm (NSR), Atrial Fibrillation (AF), First-Degree Atrioventricular Block (IAVB), Left Bundle Branch Block (LBBB), Right Bundle Branch Block (RBBB), Premature Atrial Contraction (PAC), Premature Ventricular Contraction

(PVC), ST-Segment Depression (STD), and ST-Segment Elevation (STE).

## 2.2 Data processing and compression

The data processing technique for CPSC2018 in ECG analysis entails crucial steps aimed at transforming and standardizing the electrocardiogram data. Initially, the sampling frequency of the original data is reduced by a factor of 10, achieving a 50Hz rate, which decreases data size while preserving essential ECG details. Subsequently, for each data point, 1500 sequential potential values were sampled from the full signal, using zero padding if the signal has less than 1500 values and excess values been dropped, to maintain consistency in segment length across various recordings. This is followed by a simple normalization process that standardizes the potential values, ensuring that amplitude variations in the ECG signals are uniform and comparable. Lastly, all leads, were stacked side by side to replicate the conventional paper-printed ECG layout, facilitating visual inspection and analysis.

## 2.3 Model architecture

In this study, we introduce a novel neural network architecture specifically designed for advanced ECG feature extraction and classification tasks. The architecture commences with an input layer accommodating the predefined input shape, followed by a series of convolutional layers with varying filter sizes and strides, each accompanied by layer normalization and then the aSoftmax activation. Notably, the model integrates custom global max and min pooling layers after convolutional stages, effectively capturing extremal features. These pooled outputs are subsequently concatenated and flattened, leading to a final classification layer implemented to classify inputs into 9 different types of arrhythmias. Table 2 and Figure 2 captures the summary of the model architecture and flow of proposed methodology. Further in the paper we have discussed about kernel weight sharing and aSoftmax activation, which are cornerstone to the model's performance.

**Table 2**

Model Body Architecture

| Type / Stride | Filter Shape | Output Shape |
| --- | --- | --- |
| Input Signal | - | 12 x 1500 x 1 |
| ConvB 1x3 / s1 | (1 x 3 x 1) x64 | 12 x 1500 x 64 |
| ConvB 1x3 / s1 | (1 x 3 x 64) x128 | 12 x 1500 x 128 |
| ConvB 1x3 / s2 | (1 x 3 x 128) x128 | 12 x 750 x 128 |
| ConvB 1x9 / s1 | (1 x 9 x 128) x128 | 12 x 1500 x 128 |
| MinMaxPool | - | 12 x 256 |
| MinMaxPool | - | 12 x 256 |
| Concat | - | 1 x (6144) |
| SoftMax Classifier | 6144x9 | 1x9 |

*s1*: (1, 1) stride; *s2*: (1, 2) stride; *MinMaxPool*: Lead-wise min and max pooling; *ConvB*: Convolution Block for with Conv2D, LayerNormalization, and aSoftmax Activation

## 2.4 Kernel weight sharing

Kernel weight sharing is a concept in Convolutional Neural Networks (CNNs), particularly when dealing with multi-dimensional data like electrocardiogram (ECG) recordings. In the context of ECG data, which typically consists of multiple leads, kernel weight sharing can be applied in a unique way. When dealing with multi-lead ECG data, each lead can be thought of as a separate channel (like RGB channels in images). However, in our study, the leads are not stacked as channels but treated as separate dimensions. And the same convolutional kernel can be applied across different leads, sharing weights across these. This approach can capture the inter-relationships between different leads while maintaining the low trainable weights.

Assume we have an ECG signal with $L$ leads and a convolutional kernel $K$. The convolution operation for a single lead $l$ at position $(i, j)$ in the ECG data can be represented as:

$$(K * ECG_l)(i, j) = \sum_m \sum_n K(m, n) \cdot ECG_l(i - m, j - n)$$

where $ECG_l$ represents the $l^{th}$ lead of the ECG signal, $K$ is the shared kernel, and $(i, j)$ are the spatial coordinates in the signal. In the case of kernel weight sharing across different leads, this operation is applied independently to each lead, but with the same kernel $K$. This means that for each lead $l$, the same kernel $K$ processes the signal, enabling the network to learn features that are dominating across the different leads. This approach is particularly useful for ECG data, as different leads may capture different aspects of the heart's electrical activity, but the underlying features (like QRS complexes, T-waves, etc.) have a consistent construct and nature of occurrence across leads. By sharing weights across leads, the CNN can efficiently learn these consistent features, improving its ability to generalize and reducing the risk of overfitting.

### 2.5 aSoftmax activation

The activation function described by Eq. X which I'll refer to as "Absolute Softmax Activation" (aSoftmax), introduces a novel approach to modifying input tensors for use in the proposed network. This function operates by first taking the absolute value of each element in the input tensor, ensuring that all inputs are non-negative. This step is crucial, especially in ECG signal processing, where the polarity of signal components varies, and absolute magnitudes are more informative than raw values.

$$A(x) = x \odot \text{softmax}(|x|)$$

Subsequently, the SoftMax function is applied to these absolute values. The SoftMax transformation is widely used in neural networks for its ability to convert a vector of values into a probability distribution, where the relative scale of each element is preserved, but all values are transformed to lie in the range (0,1) and sum to 1. The relativistic amplification/suppression constant then multiplied with original signal to enhance the underlying ECG feature. By combining these two steps, the aSoftmax activation ensures that the relative importance of each input feature is maintained and normalized in a probabilistic framework. This is especially important in ECG classification, where the strength and duration of various signal components (like QRS complexes, P-waves, etc.) are key indicators of cardiac health. Standard activation functions like ReLU might not handle such variations or will require more training to converge effectively, leading to loss of crucial information. Figure represents the plot of aSoftmax activation and first order derivate of it.

### 2.6 Loss and optimizer

Due to the noisy nature of ECG signal we have adopted Adamax for optimization of the proposed model. Adamax is a variant of the Adam optimizer [23], which is itself an extension of stochastic gradient descent [24] that has been widely adopted in deep learning. The Adamax optimizer is defined by the following equations:

First moment (mean) estimate: $m_t = \beta_1 \cdot m_{t-1} + (1 - \beta_1) \cdot g_t$

Second moment (uncentered variance) estimate: $u_t = \max(\beta_2 \cdot u_{t-1}, |g_t|)$

Here, $g_t$ is the gradient at time step $t$, $m_t$ and $u_t$ r are the estimates of the first and second moments respectively. The parameters $\beta_1$ and $\beta_2$ are exponential decay rates for these moment estimates, typically close to 1.

$$\theta_{t+1} = \theta_t - \frac{\alpha}{u_t + \epsilon} \cdot m_t$$

In the parameter update equation, $\theta_t$ represents the parameters at time step $t$, $\alpha$ is the learning rate, and $\epsilon$ is a small constant to prevent division by zero.

Categorical Focal Cross entropy loss [25] is adopted as loss function which is a variant of the standard cross-entropy loss function, which is commonly used for classification tasks. This loss function is designed to address class imbalance by down weighting the loss assigned to well-classified examples.

The Categorical Focal Cross entropy loss is defined as:

$$L = -\sum_{c=1}^{M} y_{o,c}(1 - p_{o,c})^\gamma \log(p_{o,c})$$

Here:

$M$ is the number of classes.

$y_{o,c}$ is a binary indicator of whether class $c$ is the correct classification for observation $o$.

$p_{o,c}$ is the predicted probability of observation $o$ being of class $c$.

$\gamma$ is the focusing parameter, which smoothly adjusts the rate at which easy examples are downweighted. When $\gamma = 0$, Categorical Focal Crossentropy loss is equivalent to standard cross-entropy loss.

The key component here is the term $(1 - p_{o,c})^\gamma$. For correctly classified examples where $p_{o,c}$ is high, this term is close to 0, which reduces the loss contribution from these examples. For misclassified examples where $p_{o,c}$ is low, this term is closer to 1, maintaining a higher loss contribution. This mechanism helps in focusing the model's training on hard, misclassified examples.

## 2.7 Experiment setup

The CPSC2018 dataset records were stratified and split into training and testing sets. The labels from the dataset were one-hot encoded to transform the problem into a multi-class classification task. The implementation of the model was carried out using TensorFlow, and a P100 GPU was utilized for training purposes. The training of the proposed system was optimized from scratch using the Adamax optimizer. This optimizer was chosen for its effectiveness in handling sparse gradients on noisy problems. An exponential decaying learning rate was employed, starting with an initial value of 0.007, to fine-tune the model training dynamically. Additionally, categorical Focal cross entropy with label smoothing of 0.3 was used. For the evaluation of the model's performance, standard metrics such as precision, recall, F1-score, and the Area Under the Curve (AUC) were used. These metrics are critical in assessing the effectiveness of the model, especially in a multi-class classification scenario where the balance between sensitivity and specificity is crucial.

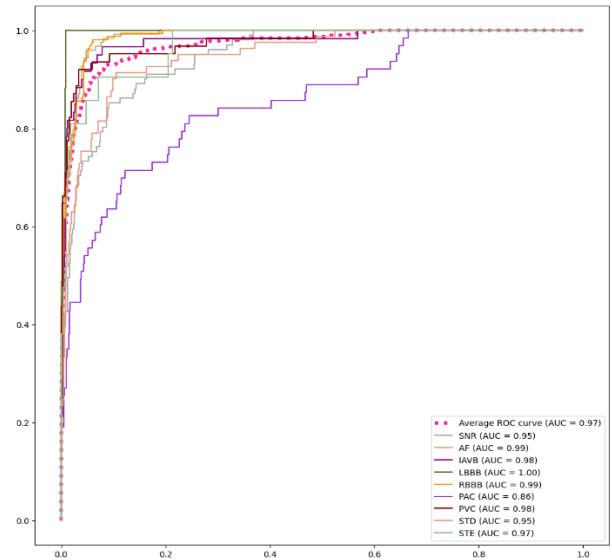

**Figure 3:** ROC-AUC curve of 9 different arrhythmia classes.

$$\underline{Accuracy:} \quad \frac{TP + TN}{TP + TN + FP + FN} \quad (25)$$

$$\underline{Recall:} \quad \frac{TP}{TP + FN} \quad (26)$$

$$\underline{Precision:} \quad \frac{TP}{TP + FP} \quad (27)$$

$$\underline{F1\text{-}Score:} \quad \frac{2 \times \text{Precision} \times \text{Recall}}{\text{Precision} + \text{Recall}} \quad (28)$$

where *TP*, *TN*, *FP*, and *FN* denote the number of true positives, true negatives, false positives, and false negatives, respectively.

## 3. Results

The evaluation of the model's performance in classifying cardiac arrhythmias was thoroughly presented in our study, classification efficacy was provided by the Receiver Operating Characteristic (ROC) curves and the Area Under the Curve (AUC) values, as illustrated in Figure 3. The model achieved a micro-average AUC of 0.97, precision 82%, recall 80%, and f1-score 81%, demonstrating its robustness in arrhythmia classification. Notably, performance varied across categories, with Atrial Fibrillation (AF), Left Bundle Branch Block (LBBB) and Right Bundle Branch Block (RBBB) showing high AUCs,

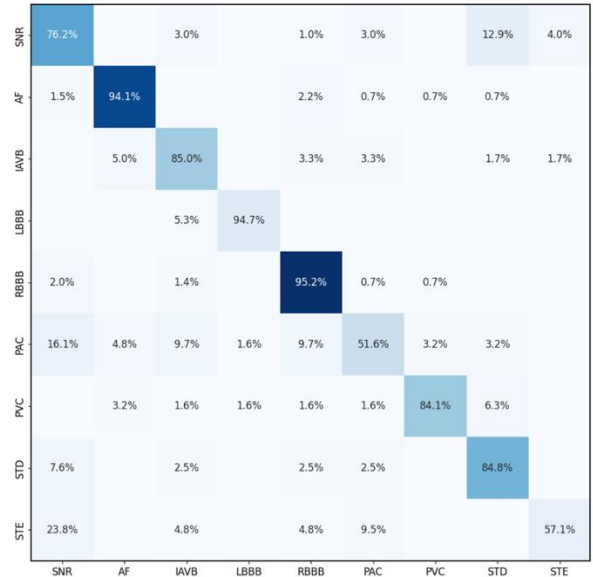

**Figure 4:** Confusion matrix of results

while categories like Intra-Atrial Ventricular Block (IAVB), Premature Ventricular

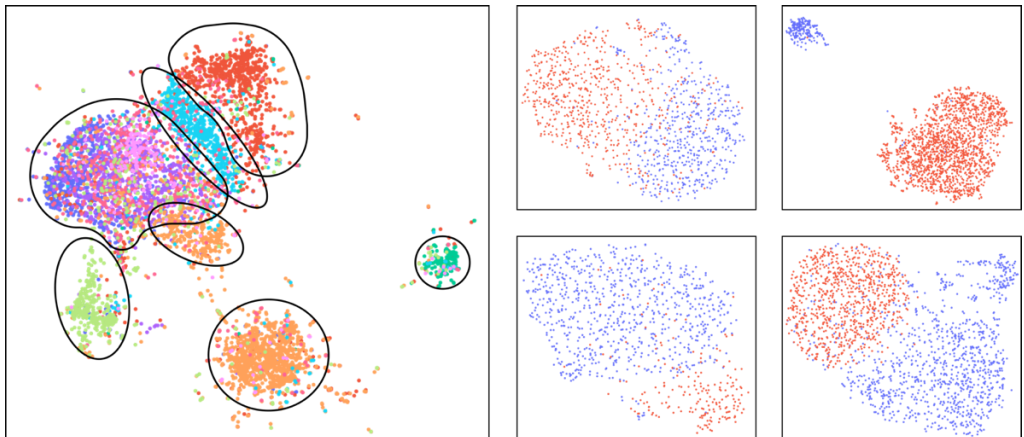

**Figure 5:** UMAP analysis of feature mapB

Contraction (PVC), ST-segment elevated (STE), and ST Depression (STD) exhibited AUCs of 0.95 and above, while challenges were observed in one category which is Premature atrial contraction (PAC) where model has achieved AUC of 0.86. The confusion matrix and classification report, as shown in Figure 4 and Table 3 respectively further elucidated the model's classification capabilities. It highlighted the model's strengths and weaknesses in identifying different arrhythmia types.

**Table 3**

AmplinetECG12 results on CPSC2018 dataset

| Class | Precision | Recall | F1 score |
|---|---|---|---|
| **NSR** | 0.7476 | 0.7624 | 0.7549 |
| **AF** | 0.9407 | 0.9407 | 0.9407 |
| **IAVB** | 0.7612 | 0.85 | 0.8031 |
| **LBBB** | 0.9000 | 0.9474 | 0.9231 |
| **RBBB** | 0.8974 | 0.9524 | 0.9241 |
| **PAC** | 0.7273 | 0.5161 | 0.6038 |
| **PVC** | 0.9298 | 0.8413 | 0.8833 |
| **STD** | 0.7614 | 0.8481 | 0.8024 |
| **STE** | 0.7059 | 0.5714 | 0.6316 |
| **Average** | 0.8190 | 0.8033 | 0.8074 |

### 3.1 Comparative Analysis

In our study, we benchmarked our model against several leading ECG classification methods, including 1D-ResNet34 [20], 1D-SEResNet34 [26], TI-ResNet18 [27], InceptionTime [28], ECGNet [28], and lightX3ECG [46]. These methods are recognized for their efficiency and smaller nature and follows similar experimental setup as this study. The comparison focused on key performance metrics such as F1 scores, computational complexity, and model compactness, as detailed in Table 4 of our study.

**Table 4**

Comparision Table

| Method | F1 on CPSC-2018 | No. Params (M) | No. FLOPs (B) | Size (MB) |
|---|---|---|---|---|
| 1D-ResNet34 [20] | 0.7684 | 16.61 | 5.91 | 58.18 |
| 1D-SEResNet34 [26] | 0.7845 | 16.76 | 5.91 | 58.75 |
| TI-ResNet18 [27] | 0.7872 | 11.39 | 1.42 | 40.51 |
| InceptionTime [28] | 0.7352 | 0.45 | 2.29 | 1.63 |
| ECGNet | 0.7880 | 1.03 | 1.97 | 3.75 |
| LightX3ECG | 0.8004 | 5.31 | 1.34 | 6.52 |
| **Ours** | **0.8070** | **0.28** | **0.20** | **1.01** |

Our model demonstrated superior performance over these established methods, while having the lowest computational cost, measured in Floating Point Operations Per Second (FLOPs), with a remarkable count of 0.76 billion. Additionally, our model excelled in terms of storage requirements. It occupied only 1.2 MB on disk, which is significantly lower than the storage requirements of the compared models.

### 3.2 Explainability

A comprehensive validation was conducted to demonstrate the interpretability and explainability of our model. This validation process involved an in-depth feature map analysis, which provided crucial insights into how our model processed and interpreted the ECG data. Additionally, the application of Shapley Values played a pivotal role in quantifying the contribution of each feature to the model's predictions. This approach was instrumental in unraveling the complex decision-making process of our deep learning model, ensuring a transparent and understandable AI system. Our findings highlighted the model's capability to classify ECG signals effectively and reliably, paving the way for enhanced diagnostic accuracy in clinical settings. The use of these advanced interpretability techniques underscored the potential of AI in medical diagnostics, providing a promising outlook for future developments in the field.

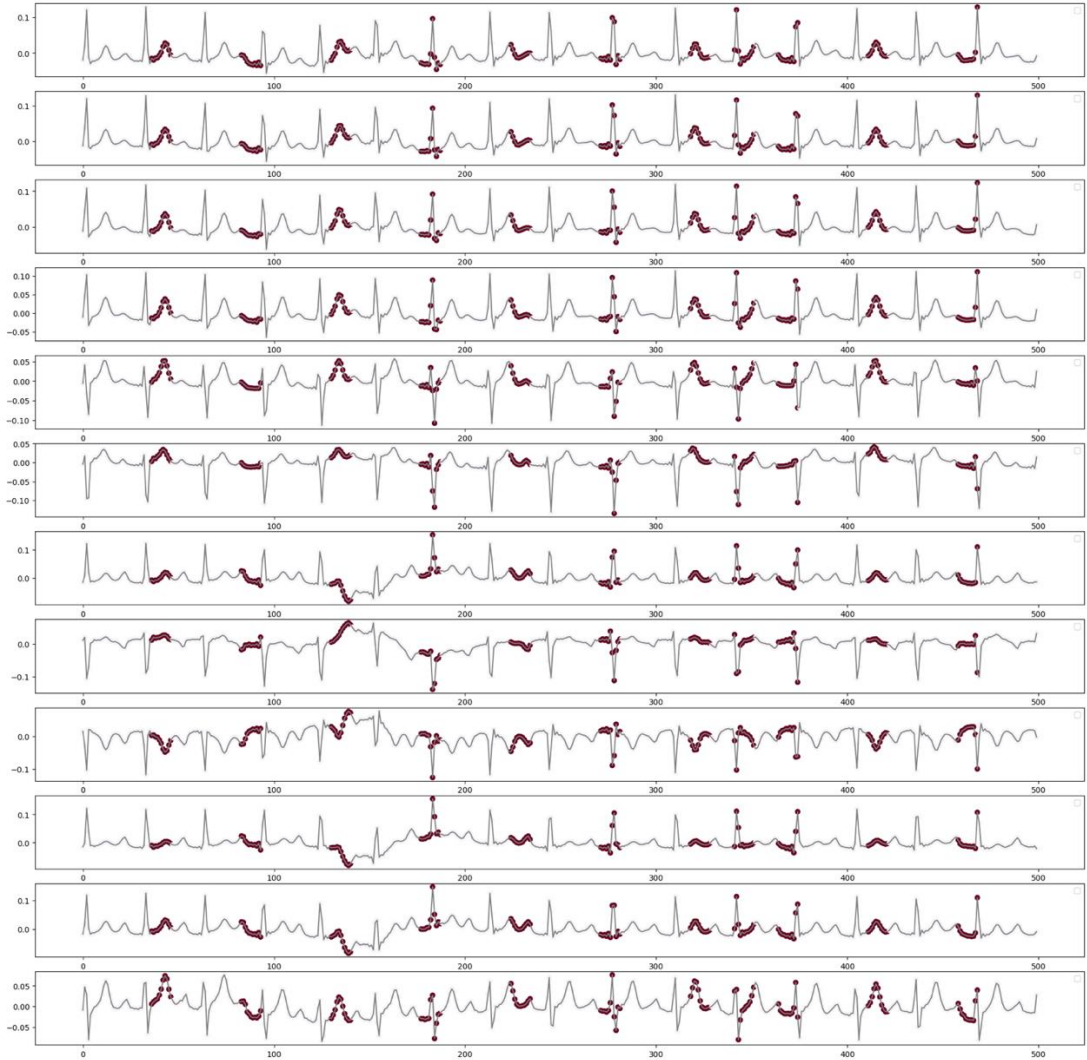

**Figure 6:** Shap value analysis for different Arrhythmia classes.

## 4. Discussion

Our system demonstrates competitive performance in diagnosis and interpretation when compared to previous works in ECG signal analysis. The improvements can be attributed to several innovative approaches: (i) We employ a unique strategy of processing three input ECG leads independently using three distinct backbones, each optimized for ECG signal efficiency. (ii) The introduction of the Lead-wise Attention module is a crucial element, significantly enhancing the system's overall performance. Furthermore, this module allows for a lead-wise explanation of the predictions, leveraging Explainable AI (XAI) techniques to identify the most influential ECG lead. However, our system has certain limitations: (i) Relying on reduced-lead ECG data might lead to missed critical information, potentially hindering the detection of specific cardiovascular abnormalities. (ii) The multi-input architecture of … poses challenges in training with small-scale datasets and results in higher storage costs, necessitating the use of practical solutions like weight pruning to mitigate these issues.

## 5. Limitation and future prospects

Our approach, which utilizes relativistic amplitude amplification, has demonstrated commendable performance in ECG signal analysis, particularly notable for its efficient use of fewer model parameters while still delivering effective predictions. This aspect of our model underscores its potential for applications where computational resources are limited. However, it's important to note that our model exhibits certain limitations in its current form, especially in comprehensively understanding rhythmic anomalies such as RR interval variation. This suggests a gap in the model's capability to fully interpret complex temporal features in ECG signals, which are critical for diagnosing certain types of cardiac irregularities. The model's current architecture, while optimized for specific aspects of ECG analysis, may require further refinement or the integration of additional techniques to enhance its proficiency in capturing and analyzing these subtle yet clinically significant variations in heart rhythm.

## 6. Conclusion

In this research article, we introduce an innovative deep learning architecture that leverages a comprehensive analysis of 12-lead, long-term electrocardiograms (ECGs) in their most condensed form to accurately classify different types of Arrhythmias. Our architecture is distinguished by two pioneering approaches: a novel activation function specifically designed to enhance the amplitude of key morphological features in the ECG signal, and a unique kernel weight sharing mechanism that effectively captures global, cross-lead ECG characteristics. This dual strategy enables our model to maintain a minimal number of trainable parameters while delivering performance that surpasses existing models in the lightweight category, as demonstrated in the CPSC2018 challenge. Quantitatively, our proposed architecture has achieved an impressive average F1-Score of 80.74% and an ROC-AUC of 97.0%, positioning it as a leading solution in the domain of Arrhythmia detection. Beyond its numerical achievements, our model marks a significant advance in the interpretability and explainability of deep learning models for ECG analysis. It exhibits a remarkable ability to identify critical ECG regions that contribute to Arrhythmia classification, a feature attributed to the innovative aSoftmax activation function. This architecture not only excels in accuracy and efficiency but also represents a significant step forward in the development of explainable AI (xAI) models. Such advancements are crucial for clinical decision-making tools, where understanding the rationale behind model predictions is essential for gaining the trust of medical professionals and facilitating their adoption in clinical settings.